\documentclass[aip,twocolumn,reprint,amsmath,amsart, amssymb,jap, floatfix]{revtex4-1}
\usepackage{natbib}
\usepackage{graphicx}
\usepackage{bm}
\usepackage{ulem} 
\usepackage{float}
\usepackage[utf8]{inputenc}
\begin{document}
\author{Arup Chakraborty}
\email{arupchakraborty719@gmail.com}
\affiliation{School of Physical Sciences, Indian Association for the Cultivation of Science, Jadavpur, Kolkata-700032, India}
\author{Bidisa Das}
\thanks{Present Address: Research Center for Sustainable Energy (RISE), TCG-CREST, Kolkata, India}
\email{bidisa.das@tcgcrest.org}
\affiliation{Technical Research Center \& School of Applied and Interdisciplinary Sciences, Indian Association for the Cultivation of Science, Jadavpur, Kolkata 700032, India}
\author{I. Dasgupta}
\email{sspid@iacs.res.in}
\affiliation{School of Physical Sciences, Indian Association for the Cultivation of Science, Jadavpur, Kolkata-700032, India}
\title{Unlocking the electronic, optical and transport properties of semiconductor coupled quantum dots using first principles methods}
\begin{abstract}
{Semiconductor coupled quantum dots provide a unique opportunity of tuning bandgaps by tailoring band offsets, making them ideal for photovoltaic and other applications. Here, we have studied stability, trends in the band gap, band offsets, and optical properties for a series of coupled quantum dots comprised of II-VI semiconductor using a hybrid functional method. We have shown how the quantum confinement and interfacial strain considerably affect the band gap and band offsets for these heterostructures at the nanoscale. We show that the trend in band offsets obtained from our first-principles electronic structure calculations agrees with that obtained from the method of average electrostatic potential. It is found that a common anion rule for band offset is followed for these heterostructures at the nanoscale. Further, the calculated optical absorption spectra for these coupled quantum dots reveal that absorption peaks lie in the ultra-violet (UV) region, whereas absorption edges are in the visible region. In addition to electronic and optical properties, we have also explored transport properties for two representative coupled quantum dots, either having common cations or common anions, which revealed asymmetric nature in current-voltage characteristics. Therefore these semiconductor coupled quantum dots may be useful for photovoltaic, light-emitting diode, and opto-electronic devices.}
\end{abstract}
\maketitle
\section{\label{intro}Introduction}
Semiconductor quantum dots (QDs), known as "artificial atoms", exhibits fascinating electronic and optical properties due to the effect of quantum confinement.\cite{science2021_QD, Cotta_ACC_Res}  These QDs find application in several fields such as light emitting diodes, photovoltaics, photoconductors,  photodetectors, catalysts, biomedicine and environment.\cite{Cotta_ACC_Res, qd_led_Nat_photo2013} We note in passing that beyond these applications, semiconductor colloidal QDs have the potential for quantum computation due to the quantum coherent effect in QDs.\cite{q_comput1, q_comput2} Recently, the heterostructures made up of different semiconductor quantum dots received considerable attention because they have several other degrees of freedom to control their properties compared to individual semiconductor QDs. In particular, the band offset  determining the band alignment at the interface of a heterostructure is a crucial parameter for functionalization and device modeling.\cite{Alferov, Banin_ACC_Res} The discontinuities between the valance band maxima (VBM) and the conduction band minima (CBM) of the individual components at the interface of the heterostructure create valance band offset (VBO) and conduction band offset (CBO), respectively. Depending on the band alignment, heterostructures are designated as straddled or ``type-I", staggered or ``type-II'' and broken or ``type-III". In particular, the type-II semiconductor heterostructures are useful for optoelectronic devices where the photoluminescence peak of the heterostructure can be tuned to a higher wavelength in comparison to its components.\cite{Alferov, NG_adv,JPCC_us} Further type-II heterostructures are ideal for photovoltaic applications as they offer a natural separation of electrons and holes.\cite{JPCC_us, Bawendi_JACS}
\newline
In nano-heterostructures, one can tailor the properties by tuning the sizes of its components due to the effect of quantum confinement. In addition, the properties of semiconductor heterostructures at nanoscale can also be controlled by inter-facial strain due to lattice mismatch between different components.\cite{Chem_Rev_2020} It was shown that band gap and band offsets for ZnSe/ZnTe heterostructure might be controlled by adequately controlling the epitaxial strain.\cite{Yadav} Khoo et. al.\cite{khoo} explored the combined effect of quantum confinement and strain in CdS/ZnS core/shell nano-heterostructures and found that band gap for these systems can be tuned by changing the shell thickness and it is also possible to manipulate the band alignment from type-I to type-II using the combined effect of confinement and strain. Li and Wang had also shown the  significant effects of strain and quantum confinement on the band gap and band offset of CdSe/CdS and CdSe/CdTe core-shell heterostructure.\cite{Li_wang} Shen et al. showed CdSe/ZnSe core-shell structures could be used as quantum dot light emitting diode, which provides red, green, blue light with external quantum efficiencies of 21.6\%, 22.9\%, 8.5\% . \cite{shen_nat_photo2019} Beyond these, there are several reports on II-VI semiconductor colloidal coupled dots, where the electronic and optical properties, for instance, absorbance, were tuned by controlling band offsets.\cite{Banin_ACC_Res, Banin2, Banin3, Banin4, watson_AMI_2021}
\newline
Furthermore, the semiconductor coupled dots or hetero-dimers has similarity with molecules in molecular electronics\cite{Tao_Nat_transport,ratner_trans}, where a single molecule with a donor part and an acceptor portion connected by
a bridge acts as a molecular rectifier as shown by Aviram and Ratner.\cite{ratner_trans} Molecular rectifiers made from organic molecules are generally unstable and device fabrication is difficult, while colloidal nanostructures are expected to have an advantage from device fabrication viewpoint. We recently showed this concept for a type-II CdS-ZnSe colloidal coupled dots, where band alignment at the interface plays a crucial role.\cite{transport_arxiv}
\newline
A reliable estimate of band offsets and understanding trends in band offsets in semiconductor heterostructures is an essential issue for practical applications. Density functional theory (DFT) is an efficient tool for studying the electronic structure of these heterostructures.\cite{cms2018_vanDeWalle} It is well known that the standard {\it ab-initio} electronic structure calculations based on DFT within the local density approximation (LDA) or generalized gradient approximation (GGA) underestimate the band gap and therefore band offsets. To overcome this problem, hybrid functionals, where exact Hartree-Fock exchange is mixed with local or semi-local exchange correlation functional, are a better option and it offers a reasonably good agreement with the experimental band gap and therefore band offsets.\cite{Wadhera,cms2018_vanDeWalle} Since the percentage of Hartree-Fock(HF) exchange depends on the systems, it is vital to find the suitable fraction($\alpha$) of HF exact exchange so that the band gap of the system is in agreement with the experimental value.
Recently, Huang et. al. studied nitride-based semiconductor heterostucture and predicted band gap and band offsets using machine learning model with screened hybrid functional (HSE) and DFT-GGA.\cite{ML_JMCC2019} Oba and Kumagai recently reviewed many semiconducting materials and their heterojunction based on first-principles calculations. \cite{Oba_review} They performed a comparative study of different first principles methods and properties such as band gap, effect of defect, optical properties etc.
While there are many reports\cite{Zunger,Zunger2,Zunger3,Passquarello,Passquarello2,Passquarello3,Wei,Wadhera,cms2018_vanDeWalle,PRB_III_V_2001, PRB2016_Leor, ACS2014, JMCC2020, PCCP2020} to understand the trends in band-offsets in II-VI and III-V bulk semiconductor heterostructures, there is hardly any systematic studies of trends in band offsets for heterostructures at the nanoscale.\\
In this paper, we have systematically studied the stability, band gap, band offsets and optical properties of a series of common anion and common cation-based II-VI semiconductor heterostructures at the nanoscale using a hybrid functional method. Our systems consist of dimers of a stable cage-like $A_{12}B_{12}$ and $M_{12}N_{12}$ clusters, where we considered coalescence of all possible dimer interactions. The band offsets are calculated from band decomposed charge density plot as suggested by N. Ganguli {\it et. al.}\cite{NG_PRB} Further, we have compared our results on band offset with a complementary approach proposed by Hinuma {\it et. al.}\cite{Hinuma} derived from the average electrostatic potential. Our calculations reveal that both techniques provide similar trends for the calculated band offsets. Next, we have calculated the optical properties of these coupled quantum dots and discussed the trends.
We recently explored the transport properties of type-II heterostructure (CdS/ZnSe) with different cations and anions.\cite{transport_arxiv} It was observed that it provides asymmetric I-V curves, which is a characteristic of a diode or switching device. So it would be interesting to explore the transport property of these heterostructures having common cations or common anions. Following our previous study\cite{transport_arxiv} on transport properties, we also have calculated the transport properties for two representative heterostructures, one with common cation and other with common anion system.
\newline
The remainder of this paper is organized as follows. In section~\ref{method_cqd}, we describe our theoretical methods and structural details. Section ~\ref{result_cqd} is devoted to results and discussion on electronic structure, optical and transport properties of coupled quantum dots. Finally, we present a summary and conclusions in section~\ref{summary}.
\section{\label{method_cqd}Computational details}
All the electronic structure calculations presented in this work are carried out using the Vienna {\it Ab-initio} simulation package (VASP).\cite{vasp1, vasp2} Our calculations are performed using the generalized gradient approximation (GGA) of Perdew-Burke-Ernzerhof (PBE) functional,\cite{pbe} and Heyd-Scuseria-Ernzerhof (HSE06),\cite{hybrid1, hybrid2} hybrid functional for exchange-correlation potential. The fraction of HF exact exchange ($\alpha$) is chosen to be 0.25 in the HSE06 hybrid functional.  
The electron-ion interactions are described by the projector augmented wave (PAW)\cite{paw} potentials, as implemented in the VASP code. The wavefunctions were expanded with plane waves with a kinetic energy cutoff of 500 eV. A sufficient vacuum (10 \AA) is maintained in all three directions to get rid of the interaction between periodic images of these nano-heterostructures. 
A single point $\Gamma$-centered k-mesh was used to sample the Brillouin zone. The density of states (DOS) was evaluated by the Gaussian smearing method with the Gaussian width of 0.01 eV. The ionic positions were optimized until the absolute value of the interatomic forces became less than 0.01~eV/\AA.
\begin{figure} 
\centering
\includegraphics[scale=0.3]{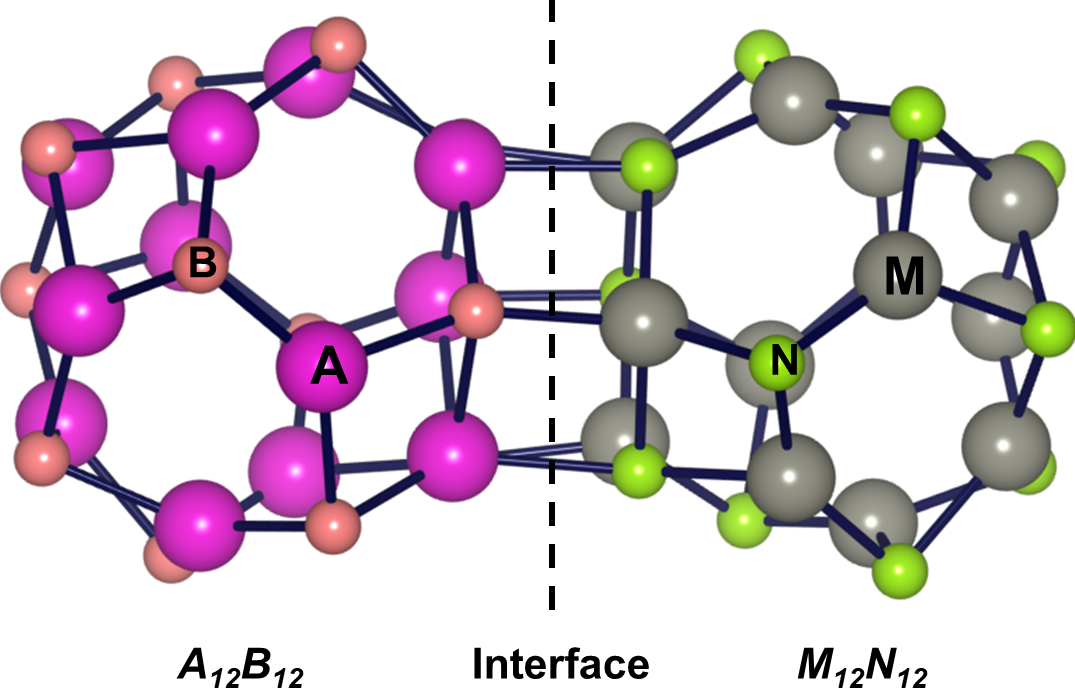}
\caption{Schematic representation of the optimized heterostructure.}
\label{struc_sml}
\end{figure}
\newline
To construct an interface, we have attached two different semiconductor nanoclusters of similar sizes (see Fig.~\ref{struc_sml}). These individual nanoclusters have a Fullerene-like cage structure with 12 cations and 12 anions. These kinds of small magic sized cluster of Zn$_{12}$O$_{12}$\cite{NG_JAP}, Cd$_{12}$S$_{12}$\cite{SG}, Zn$_{12}$S$_{12}$\cite{chen}, In$_{12}$As$_{12}$\cite{In12As12}, M$_{12}$N$_{12}$(M=Al,Ga)\cite{assem1} etc, have been reported by different groups. These clusters consist of six members, four members and two members rings. There are several possibilities of dimer interaction.\cite{assem1,assem2,assem3}
We note that our previous study revealed that the configuration having six bonds at the interface out of other possible configurations is energetically favorable, as shown in Fig.~\ref{struc_sml}.\cite{transport_arxiv} So, we only focus on that particular lower energy configuration (Fig.~\ref{struc_sml}) for this study.
\newline
To model the electronic transport, the coupled dots is placed between two gold electrodes, Au(111), forming a stable two-probe junction. The electronic transport through the coupled dots in the Au(111)$-$coupled dots$-$Au(111) based two probe junction is calculated using a combination of non-equilibrium Green’s Function (NEGF) formalism and DFT as implemented in Quantumwise software. \cite{ATK1,ATK2} The transport studies consist of two Au(111) electrodes on both sides, with a coupled dot connected in between. The scattering region consists of a coupled dot and a portion of the electrodes containing six 5x5 layers of gold atoms (see Figure 7 for model). Polarized double zeta (DZP) basis-set was used for all atoms except Au (SZP basis-set was used) with GGA-PBE \cite{pbe,Zunger-pbe} functional for the exchange correlation method. We have noted that we have performed the calculations of transport properties with GGA-PBE only to save computational cost.
The transmission spectrum reported is calculated using the following expression: $T(E,V)=Tr[\Gamma_L(E,V)G(E,V)\Gamma_R(E,V)G^\dagger(E,V)]$, where  $\Gamma_{L/R}$ stands for the coupling matrix between two electrodes and the scattering region, $G(E,V)$ is the retarded Green's function of scattering region.   For all cases, the transmission spectra with 6$\times$6 k-points are calculated and analyzed. The electronic eigenstates of the coupled dot in the two-probe environment are denoted by Molecular Projected Self-consistent Hamiltonian (MPSH) states. The MPSH states, which can conduct owing to their overlapping molecular orbitals and align well with the electrode Fermi energy, show peaks in the transmission spectrum.
The non-linear current at an applied bias voltage $V_b$ through the contact is calculated using Landauer formula,
$I(V_b)=G_0\int\limits_{\mu_L} ^{\mu_R}T(E,V_b) dE$
\noindent where $G_{0} = {2{e^{2}}/h}$ is the quantum unit of conductance and $\mu_{L/R}$ are the electrochemical potentials of the left and right electrodes and $h$ is Planck's constant. The zero-bias conductance is given by transmission at the Fermi energy of the electrodes.  In the two-probe setup, the coupled dots states align with the Fermi energy $(E\_{F})$ of Au(111) and  the energies are reported relative to $E\_{F}$in the transmission spectra. The transmission eigenvalues are obtained by diagonalizing the transmission matrix. The number of eigenvalues indicates the number of individual channels going through the scatterer; here the coupled dot. The eigenvalues represent the strength of each channel. The eigenvalues are the actual transmission probabilities, thus ranging between 0 to 1. Most of the eigenvalues are negligibly small and only the large eigenvalues of each energy are the most important ones. If several channels are available at a particular energy, their sum and hence the transmission coefficient at this energy, may however be larger than 1.\cite{soren}
\section{\label{result_cqd}Results and Discussion}
\begin{figure}
\centering
\includegraphics[scale=0.35]{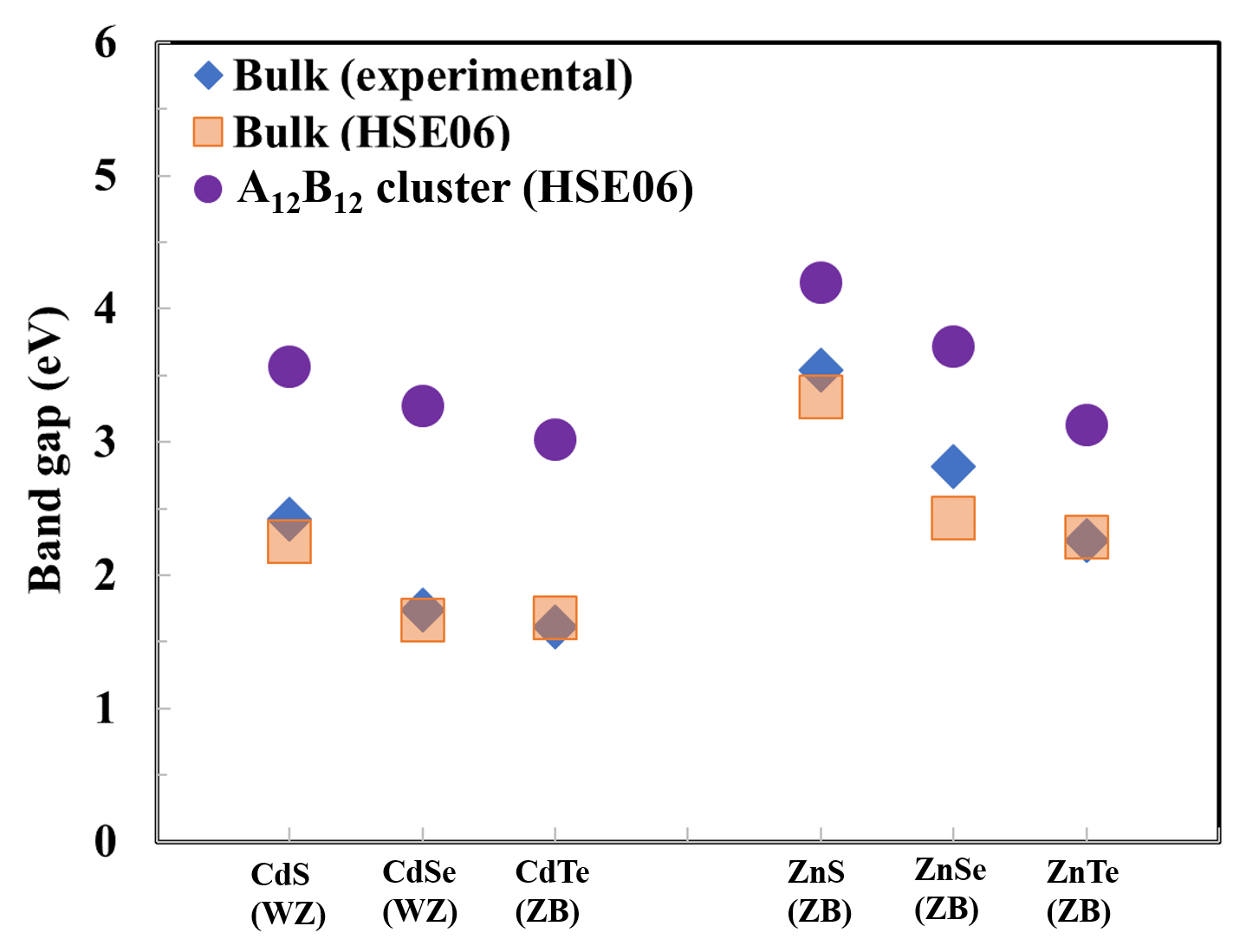}
\caption{Calculated band gaps for bulk and its nano counterparts using HSE06. The experimental band gaps for bulk systems are also given.}
\label{bnd_gap}
\end{figure}
\begin{figure}
\centering
\includegraphics[scale=0.6]{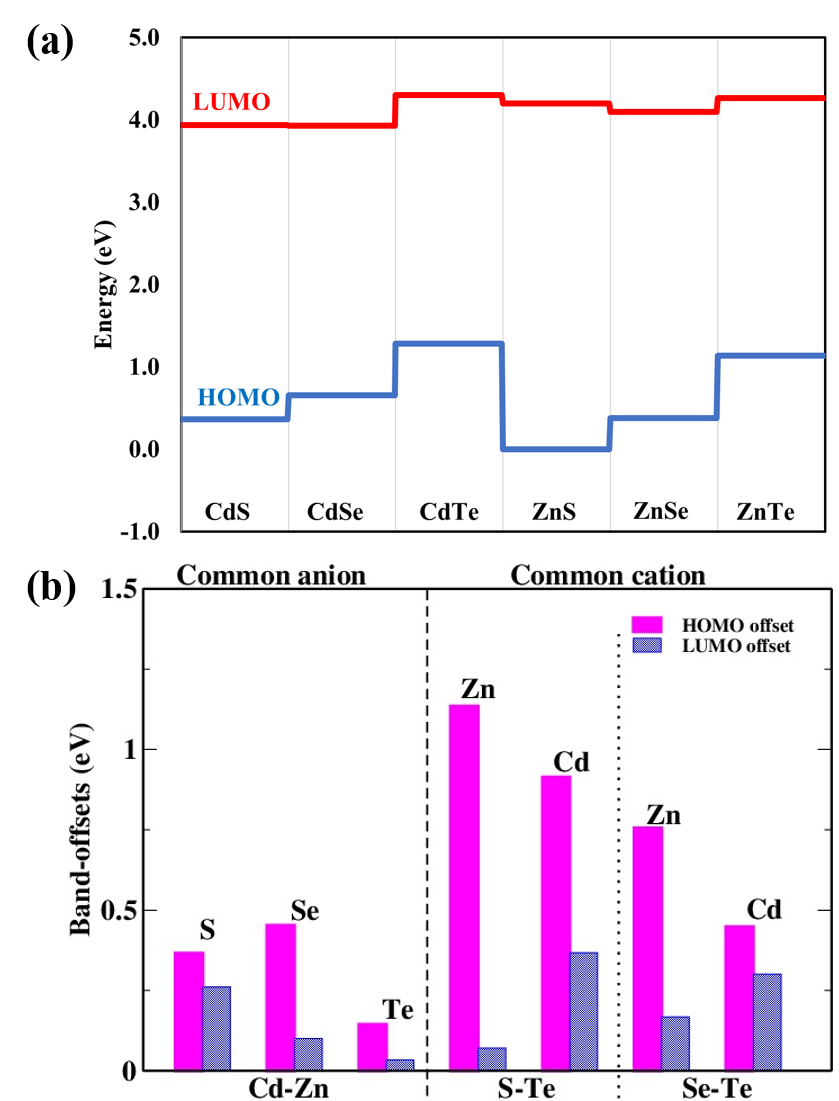}
\caption{(a) Band alignment between different nanoclusters before attachment. (b) calculated band offset from the alignment as shown in (a).}
\label{bnd_alignSm}
\end{figure}

To begin with, we have calculated the band gap of a few bulk semiconductors (CdS, CdSe, CdTe, ZnS, ZnSe, ZnTe) using HSE06 with $\alpha$ = 0.25 and the calculated results along with experimental band gap\cite{wiki} are shown in Fig.  \ref{bnd_gap}. It is observed that while the band gaps using GGA are severely underestimated, the band gaps calculated using HSE06 are in good agreement with the experimental values. For further studies, we have used HSE06 with $\alpha$ = 0.25 for our calculations, except for the transport calculations where we have used  GGA-PBE method.\\
Next, we have calculated the band gap of individual $A_{12}B_{12}$ nanoclusters of cadmium and zinc chalcogenides of similar size containing 12 cation and 12 anion atoms. The results of our calculations are presented in Fig.~\ref{bnd_gap}. It is to be noted that the values of the band gap of the clusters are large compared to bulk systems due to the effect of quantum confinement. Similar to bulk, the band gap decreases with an increase in the atomic number of the anion from S to Se to Te (see Fig. ~\ref{bnd_gap}).\\ 
To have an idea of the band-offsets and to determine the nature of the heterostructure, we have plotted the highest occupied molecular orbitals (HOMO) and lowest unoccupied molecular orbitals (LUMO) for individual nanoclusters keeping the vacuum level aligned. From the position of HOMO and LUMO states for two different semiconductors, we can infer the  nature of the heterostructures. The results of our calculation are shown in Fig.~\ref{bnd_alignSm} and for example, we find from Fig.~\ref{bnd_alignSm}(a) that the band alignment of CdS-ZnS is of type-I whereas CdSe-CdTe is of type-II.
Fig.~\ref{bnd_alignSm}(a) also provides a rough estimate of the HOMO and LUMO offsets simply from the energy difference of the HOMO and LUMO states.
The results of our calculation for HOMO and LUMO offsets using such an approximate method are shown in Fig.~\ref{bnd_alignSm}(b). We find common anion system displays small value of band offset.\\
We have discussed in the last paragraph band alignment of heterostructure from the knowledge of individual clusters. We shall now discuss our results based on self-consistent calculations. As there are several possibilities to form the heterostructure, we observed that the configuration having the highest number of bonds (six bonds) at the interface is the most favorable configuration.\cite{transport_arxiv} So, we only have considered configuration-V (6 bonds) in this paper.
We have discussed the trends in binding energies, band gaps and band offsets for some common anion and common cation systems of II-VI semiconductor heterostructures at nanoscale combining CdS, ZnS, CdSe, ZnSe, CdTe, ZnTe systems. Before discussing the trend in band offsets for common cation or anion systems we shall discuss the stability of the coupled clusters.
\subsection{ Binding energies of coupled clusters:}
The binding energy ($\delta{E}$) of the $A_{12}B_{12}$-$M_{12}N_{12}$ coupled dot can be calculated from the total energies of coupled systems and their constituent systems. It is defined as follows.
\begin{eqnarray}
\delta{E} = [E_{total}(A_{12}B_{12}/M_{12}N_{12}) &-& (E_{total}(A_{12}B_{12}) \nonumber\\ &+& E_{total}(M_{12}N_{12}))]
\nonumber
\end{eqnarray}
Here $E_{total}(A_{12}B_{12}/M_{12}N_{12})$, $E_{total}(A_{12}B_{12})$, $E_{total}(M_{12}N_{12})$ are the total energy of coupled system and total energies of $A_{12}B_{12}$ and $M_{12}N_{12}$ respectively. The binding energies for common cation systems (e.g. (Se,Te),(S,Te)) are all negative as shown in TABLE \ref{BE}, which indicates the formation of a coupled cluster is energetically favorable. 
In common anion systems (Cd$_{12}$X$_{12}$-Zn$_{12}$X$_{12}$), $\delta${E} increases with an increasing atomic number of common anion (X) as shown in TABLE \ref{BE}. It is observed that, similar to bulk\cite{Zunger2}, for the common cation system, the Cd based systems have smaller binding energies compared to Zn system (see TABLE \ref{BE}).
\begin{table}
\caption{Binding energy and band gap for different coupled clusters using GGA-PBE and HSE06. (all are in eV).}
\centering
\scalebox{0.85}{
\begin{tabular}{|c|c|c|c|c|}
\hline
Systems & $\delta{E}$ (eV)  & $\delta{E}$ (eV) & Band gap (eV) & Band gap (eV)\\ 
 & using GGA  & using HSE06 & using GGA & using HSE06  \\ 
\hline
CdZnS & -1.12 & -1.66 & 2.31 & 3.36 \\
\hline
CdZnSe & -0.63 & -1.11 & 2.07 & 3.01 \\ 
\hline
CdZnTe & -0.87 & -0.26 & 2.00 & 2.72\\ 
\hline
ZnSeTe & -0.19 & -0.41 & 2.29 & 3.08 \\ 
\hline
CdSeTe & -1.46 & -1.05 & 1.91 & 2.73 \\ 
\hline
ZnSTe & -0.16 & -0.43 & 2.22 & 3.03\\ 
\hline
CdSTe & -1.67 & -1.28 & 1.98 & 2.82\\
\hline
\end{tabular}
}
\label{BE}
\end{table}

\begin{figure}
\centering
\includegraphics[scale=0.45]{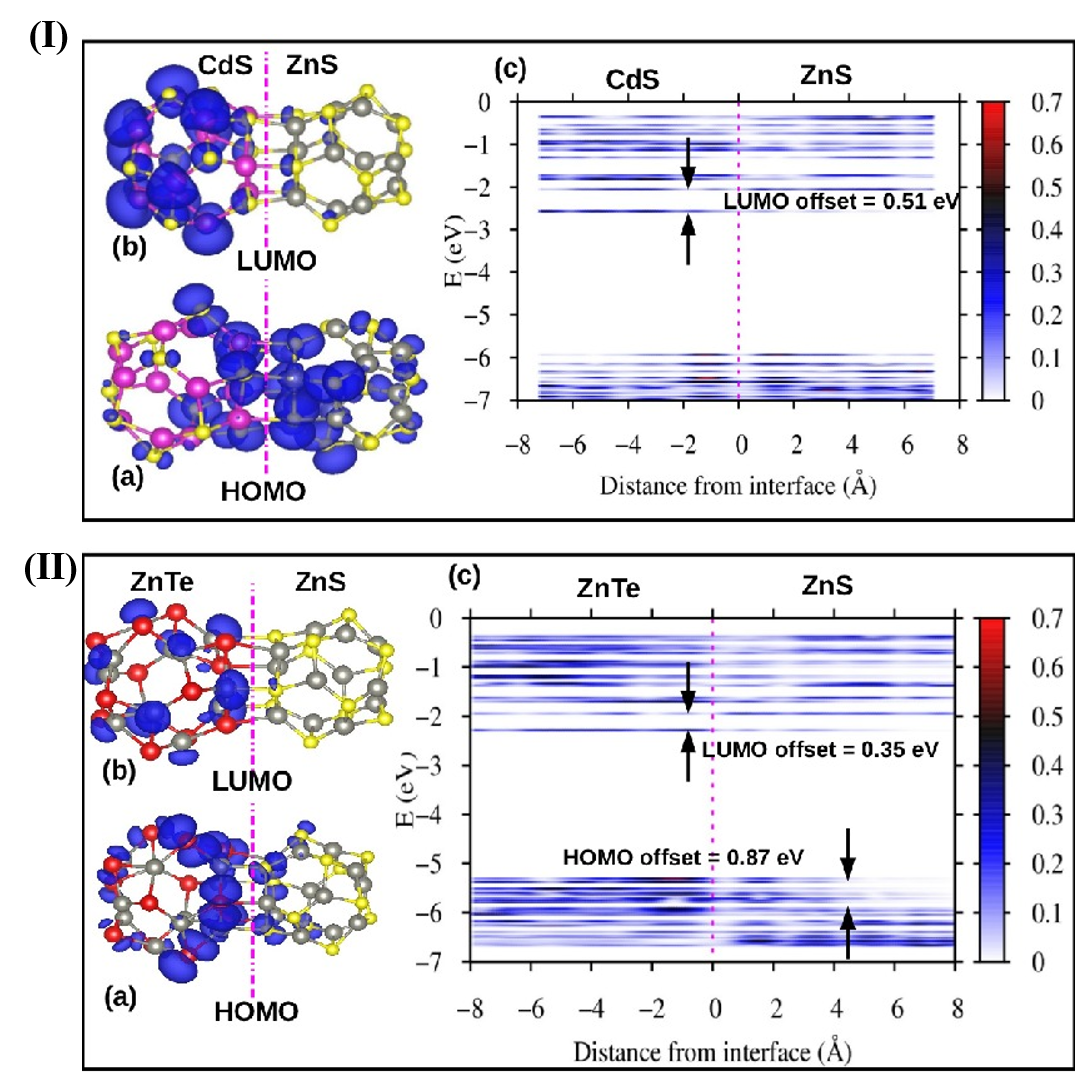}
\caption{I.(a) Iso-surface of charge density for LUMO state, (b) iso-surface of charge density for LUMO state, (c) energy resolved charge density (ERCD) for CdS-ZnS heterostructure using HSE06. II.(a) Iso-surface of charge density for LUMO state, (b) iso-surface of charge density for LUMO state, (c) energy resolved charge density (ERCD) for ZnS-ZnTe heterostructure using HSE06.}
\label{ERCD_merge}
\end{figure}
\subsection{ Band gap and band offsets:} 
\begin{figure}
\centering
\includegraphics[scale=0.35]{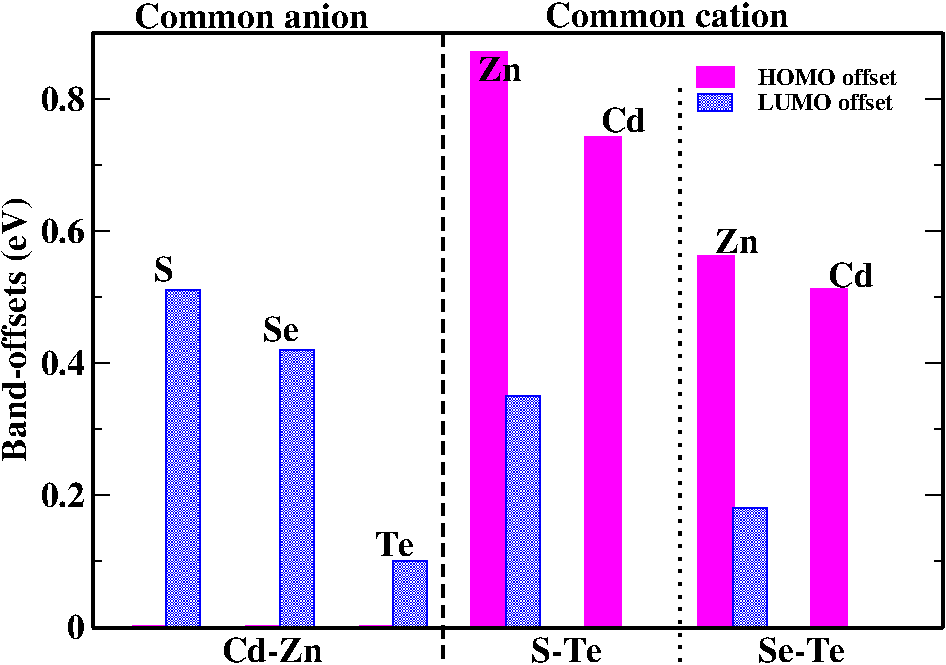}
\caption{HOMO and LUMO offsets for the coupled dots calculated using HSE06.}
\label{offset_6bnd}
\end{figure}

We have calculated the band gap for coupled clusters using hybrid functional HSE06 (see Table~\ref{BE}). An analysis of the electronic structure for the coupled dimer reveals that HOMO states have primarily contribution from anion p-states while the LUMO states have the contribution from cation s-states.\cite{NG_PRB} 
So, the HOMO offset should be dictated by the anion-p states and the LUMO offset should be dictated by cation-s states. To calculate the band-offset, we have plotted energy resolved charge density (ERCD) for a representative CdS-ZnS (common anion) coupled cluster shown in Fig.~\ref{ERCD_merge}(I). Our calculations for this common anion systems reveal the absence of HOMO offset and substantial LUMO offset of 0.51 eV. It is a quasi type-II heterostructure observed from the iso-surfaces of charge density for HOMO and LUMO as shown in Fig.~\ref{ERCD_merge}(Ia) and (Ib). ERCD for a representative common cation system, ZnS-ZnTe is displayed in Fig.~\ref{ERCD_merge}(II) and the calculated HOMO and LUMO offsets are respectively 0.87 eV and 0.35 eV. Iso-surfaces of charge density for HOMO and LUMO of ZnS-ZnTe dimer, as shown in Fig.~\ref{ERCD_merge}(IIa) and (IIb), reveal that it is a type-I heterostructure. The HOMO offsets for different II-VI semiconductor coupled dots with common anion or common cation are tabulated in TABLE ~\ref{offset}. In order to assess the impact of strain, the HOMO offsets are calculated for unrelaxed (without strain) and fully relaxed interfaces.\\
We have considered common anion systems (S or Se or Te with Cd and Zn cations). Our self-consistent calculations reveal that there are hardly any valance band offsets for the fully relaxed interfaces of common anion systems (CdZnS,CdZnSe,CdZnTe) following the common anion rule\cite{cmn_anion_book} (see TABLE~\ref{offset}). This rule is in contrast to  bulk interfaces, as observed by Wei and Zunger\cite{Zunger}. Wei and Zunger had mentioned that the presence of valance band offset in bulk semiconductor heterostructure is due to p-d coupling. In zinc blende structure with T$_d$ symmetry, both the anion-p and cation-d orbitals transform into T$_2$ representation according to the character table of T$_d$.\cite{Zunger} The A$_{12}$B$_{12}$ nanocluster has T$_h$ symmetry\cite{assem1} and in T$_h$ symmetry, anion-p and cation-d orbitals belong to two different representation (T$_g$ and T$_u$), so, it is expected that there will be hardly any p-d coupling, which is reflected in our results by the absence of HOMO offset obeying the common anion rule in the common anion systems. Symmetry is further lowered in two coupled cluster system.\cite{assem3}. 
\begin{table}
\caption{HOMO offsets for different configurations in eV using HSE06. Parenthesis in the fourth column denotes HOMO offset calculated from the method proposed by Hinuma et. al.\cite{Hinuma}.}
\centering
\scalebox{0.78}{
\begin{tabular}{|c|c|c|c|c|c|}
\hline
Systems & Bulk offset & HOMO offset & HOMO offset & LUMO offset & LUMO offset \\
 & from Ref.[~\onlinecite{Zunger}] & (Un-relaxed) & (Relaxed) & (Un-relaxed) & (Relaxed) \\
\hline
CdZnS & 0.18  & 0.0 &  0.0 (0.00) & 0.26 & 0.51 \\
\hline
CdZnSe & 0.07  & 0.0 & 0.0 (0.00) & 0.15 & 0.42\\
\hline
CdZnTe & 0.13  & 0.0 & 0.0 (0.01) & 0.17 & 0.10 \\
\hline 
ZnSeTe & 0.73  & 1.32 & 0.56 (0.36) & 0.00 & 0.18 \\
\hline
CdSeTe & 0.57  & 0.50 & 0.51 (0.29) & 0.00 & 0.00 \\
\hline
ZnSTe & 1.26  & 1.41 & 0.87 (0.67) & 0.00 & 0.35\\
\hline
CdSTe & 0.99  & 1.58 & 0.74 (0.52) & 0.67 & 0.00 \\
\hline
\end{tabular}
}
\label{offset}
\end{table}
We have further calculated HOMO offset for common cation (Cd or Zn with Se/Te, S/Te) systems. It is observed that HOMO offsets are quite large compared to common anion systems (see TABLE ~\ref{offset} and Fig. ~\ref{offset_6bnd}) because HOMO offsets are primarily dictated by two different anion-p states. 
It is observed that the HOMO offsets for common cation system decrease with an increasing atomic number of common cation as observed by Wei and Zunger\cite{Zunger} for bulk interfaces.
Further, we have calculated band offsets using a complementary approach proposed by Hinuma {\it et. al.} \cite{Hinuma} and results are displayed within the parenthesis in the fourth column of the TABLE~\ref{offset}. It suggests that the trends of HOMO offsets remain the same in both methods.
\newline
Similar to HOMO offsets, we have also calculated LUMO offset and its trends for common anion and common cation systems. 
It is observed that there is substantial LUMO offset for common anion systems since the LUMO has the contribution of cation-s orbitals and LUMO offset decreases from S to Se to Te for common anion systems. We have found that there are negligible LUMO offsets as obtained from ERCD for common cation systems compared to common anion systems. 
It is to be noted that LUMO offset obtained from the method by Hinuma et. al.\cite{Hinuma} is quite large compared to values obtained from ERCD because LUMO offset is calculated by adding HOMO offset with the difference of band gap for individual clusters and in coupled systems band gap is small compared to individual clusters. 
The trend in LUMO offset (decreasing of offset with an increase of the atomic numbers of common anion and common cations) for common anion and common cation systems obtained from two different approaches remains the same.
\newline
Further to know the effect of strain on LUMO offset for the series of small coupled dots, we have calculated the LUMO offsets for relaxed and unrelaxed structures of coupled system and results are tabulated in Table~\ref{offset}. It is observed that strain affects the LUMO offsets in these heterostructures.
\subsection{Optical properties of coupled quantum dots}
The optical properties can be calculated from the real part ($\epsilon_1$) and the imaginary part ($\epsilon_2$) of the dielectric tensor. We can calculate the absorption coefficient $\alpha(\omega)$ from the following relation:
\begin{eqnarray}
\alpha (\omega) = \sqrt{2}\omega \left[\sqrt{\epsilon_1^2(\omega) + \epsilon_2^2(\omega)} - \epsilon_1(\omega) \right] ^{1/2}.
\end{eqnarray}
\begin{figure}
\centering
\includegraphics[scale=0.55]{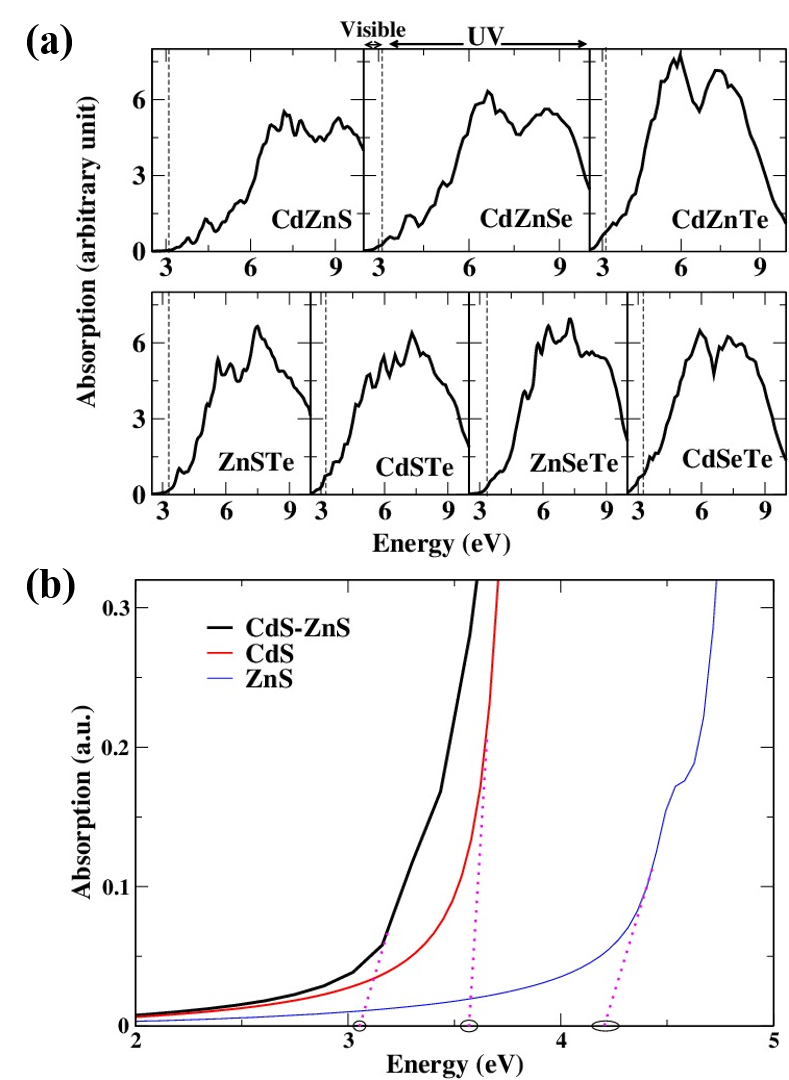}
\caption{(a) Absorption coefficient for common anion and cation systems using HSE06. (b) Absorption coefficient for the representative dimer, CdS-ZnS common anion system and its components using HSE06. Absorption edges for dimer and its individual components are pointed out by circles on the energy axis.}
\label{abr_hse}
\end{figure}
The absorption spectra is basically governed by the selection rule for optical interband transition. For nanocrystals, the optical properties can be controlled by tuning the size of the particle. The calculated absorption spectra for common anion and cation systems are displayed in Fig.~\ref{abr_hse}.
It is observed that for most of the systems, absorption edges are in the visible region, especially for common cation systems, but the absorption peaks are in the UV region.
\newline
It is to be noted that the absorption edges of individual nanocrystals are not in the visible range, but when two nanocrystals are combined to form a coupled dot, then absorption edges can be in the visible region as shown in Fig.~\ref{abr_hse}(b).
\newline
The absorption edges for common anion systems with an increasing atomic number of common anion atoms are shifted from UV to the visible region due to the reduction of band gap, but the absorption peaks remain at the UV region. On the other hand, for the common cation (Cd or Zn) systems, the absorption edges are shifted from visible to UV region as we go from Cd to Zn since the band gap changes accordingly.
\begin{figure*}
\centering
\includegraphics[scale=0.6]{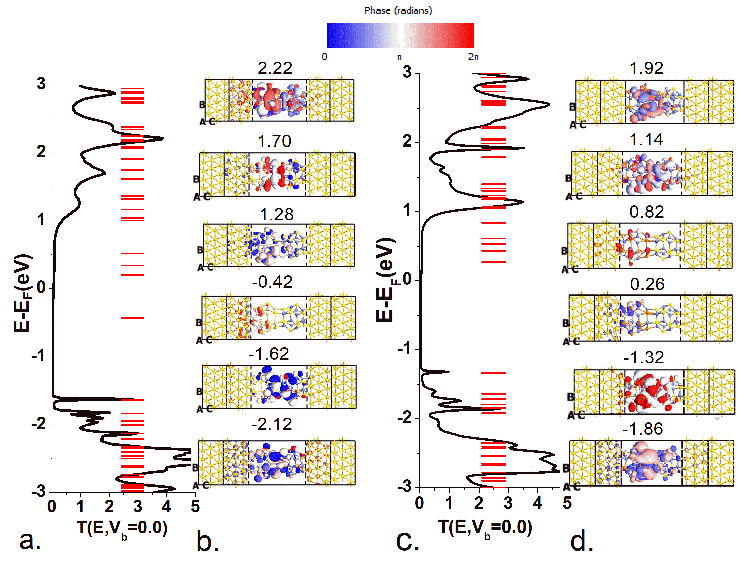}
\caption{ \textbf{(a)} Transmission spectra, $T(E)$, presented for Au(III)$-$CdS-ZnS$-$Au(III) two-probes setup considering both spins. The transmission spectra are calculated for zero applied bias. The horizontal lines shown at the right of $T(E)$ plot show the MPSH states of the coupled QD in the two probe set up.  \textbf{(b)} Transmission eigenstates/channels calculated from the $T(E)$ for CdS-ZnS coupled dot. \textbf{(c)} Transmission spectra presented for Au(III)$-$ZnS-ZnTe$-$Au(III) two-probe setup considering both spins for no applied bias. Horizontal lines shown at the right are MPSH states.  \textbf{(d)} Transmission eigenstates/channels calculated from the $T(E)$ of ZnS-ZnTe coupled dot. The phases of the eigenstates are shown in two color format.}
\label{fig1}
\end{figure*}

\begin{figure}
\centering
\includegraphics[scale=0.45]{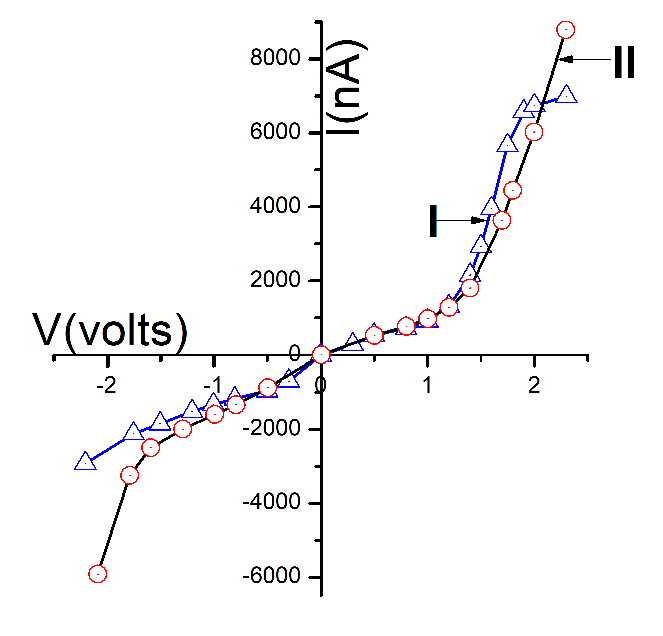}
\caption{The calculated current versus voltage (I-V) curve for CdS-ZnS (I) and ZnS-ZnTe (II) coupled dot.}
\label{fig3}
\end{figure}
\subsection{\label{transport}Transport properties} 
This section explores the transport properties of the similar size coupled dots having common anion or common cation. We have considered two representatives coupled dots, one is CdS-ZnS with common anion and other is ZnS-ZnTe with common cation. The electronic transport study of these coupled dots is done by inserting them between two Au(111) electrodes. The transmission of electrons in such a scenario is determined by the band alignment within the coupled dot heterojunction,  along with the alignment of Au Fermi energy with the coupled dot states. The systems considered here are asymmetric by construction, hence the HOMO and LUMO are localized on different sides of the coupled dot, and they interact differently with source and drain Au electrodes. The nature of the electron transport is then understood from the transmission spectra ($T(E)$) calculated at zero applied bias for the Au(111)$-$CdS-ZnS$-$Au(111) (Figure \ref{fig1}a.) and for Au(111)$-$ZnS-ZnTe$-$Au(111) (Figure \ref{fig1}c). 
Along with  $T(E)$, we also plot the MPSH eigenstates of the coupled dot, which are the states of the coupled dot modified by the presence of Au(111) electrodes (see Figure \ref{fig1}. 
It is seen that for CdS-ZnS and ZnS-ZnTe  dots, the $T(E)$ exhibited very low transmission around electrode Fermi energy, and there is a significant gap between important transmission peaks appearing for both positive and negative energies (Figure \ref{fig1}a and Figure \ref{fig1}c). While we observe several MPSH states appear at energies in between, these appear to be mostly electrode centric, and unimportant for transmission. It is seen that for CdS-ZnS dot, there is an energy gap of  ~2.7 eV between large transmission peaks.
Though the overall $T(E)$ appears similar in both the cases, the nature is different in the positive energy side for the ZnS-ZnTe dot with larger peaks compared to the CdS-ZnS dot, indicating better alignment of the LUMO  with Au Fermi energy for the former. We have plotted a few eigenstates near $E-E_F=0$ region to understand the origin of the peaks in $T(E)$. The eigenstate of the coupled CdS-ZnS dot (Figure \ref{fig1}b) contributing to the transmission peak at -1.62 eV, shows large amplitude at the central region, which can be easily related to the molecular-orbital shown in Figure \ref{ERCD_merge} Ia, which is then modified in the two-probe environment. Similarly, the  transmission eigenstate corresponding to the peak around 1.28 eV in the $T(E)$ (Figure \ref{fig1}a), has large amplitudes only on the left side, but, the eigenstates also must have large amplitude on the right side of the scattering region,to ensure a higher  probability of transmission.  So only a small peak with transmission co-efficient of 0.48, is observed here, which indicates less electronic transmission. On the other hand, the transmission eigenstate at -0.42 eV  correspond to a state which is localized, and barely has any amplitude around the right side of the scattering region, and therefore no transmission peak is observed at all around this energy. Higher transmission is observed through the eigenstates at -2.12, 1.70 and 2.22 eV as large amplitude is observed through-out the scattering region, including the heterojunction along with some amplitude near the electrodes, indicating formation of good transmission eigenchannels. The transmission function is non-zero at electrode Fermi energy $(E-E_F=0)$, and the zero-bias conductance is $0.41G_0$ ($G_0 = 7.748X10^{-5}$ S). 
In the case of ZnS-ZnTe coupled dot, we have similarly plotted a few significant transmission eigenstates (Figure \ref{fig1}d). The transmission eigenstates at 0.26 and 0.82 have small amplitudes near the electrodes. However, eigenstates at -1.32, -1.86 eV or 1.14 and 1.92 eV show large amplitudes, which are more distributed over the coupled dot as well as the electrode coupled dot interface.  Consequently, there are large transmission peaks corresponding to these energies. Zero-bias conductance calculated for ZnS-ZnTe coupled dots is $0.27G_0$. In the case of ZnS-ZnTe coupled dot the energy gap between the important transmitting states is {\it ca.} 2.4 eV with few non-transmitting states appearing in between. Due to the inherent asymmetric nature of the occupied and the unoccupied molecular orbitals of the coupled dots, it is expected to show an asymmetric current-voltage characteristics and a consequent diode-like behavior. The current-voltage (I-V) curves for the two-probe junctions based on both the coupled dots are shown in Figure \ref{fig3}. We find both the I-V curve to be asymmetric but more pronounced asymmetry is observed for CdS-ZnS coupled dot, which starts increasing rapidly after a positive bias voltage of 1.2 V and reaches a maximum around 2 V after which the current levels off. In the case of negative bias voltage, the change in current is not very pronounced till -2.3 V and the current remains to be much lower than those obtained by applying similar positive bias voltages. This may be co-related to the $T(E)$ as electronic transmission starts increasing from 0.8 eV to give a small peak around 1.28 eV and contribute to increasing current compared to the sharp negative energy peaks appearing only after -1.62 eV or lower. In the case of the I-V curve of ZnS-ZnTe coupled dot the current increases sharply after 1.5 V and current keeps the increasing trend. In contrast, the increase in current at negative applied bias is smaller, attributing to asymmetry in the I-V curve.   
\section{\label{summary}Conclusions}
We have systematically studied trends in band offsets and optical properties for a series of small coupled dots comprising II-VI semiconductors using HSE06. It is observed that, similar to bulk, the band gaps of individual clusters decrease from S to Se to Te. After coupling two nanoclusters the nature  of the heterostructures remain same but the band offsets differ markedly. This is attributed to the effect of chemical bonding and interfacial strain at the interfaces.  
The common anion systems allow hardly any HOMO offset consistent with common anion rule. This can be ascribed to the special symmetry (T$_h$) of the nanoclusters and there is hardly any p-d coupling unlike its bulk counterparts. For common cation systems there is finite HOMO offset and it decreases with an increasing atomic number of the common cations. We have also found that there is substantial LUMO offset for common anion systems and LUMO offset decreases with an increasing atomic number of common anions, similar to the case of HOMO offset. The band gap is found to decrease with an increasing atomic number of common elements for both common anion and common cation systems. It is also important to note that we have verified our results on band offsets by two different approaches and found very similar trends. We have explored the optical property of these heterostructures  and our calculations reveal that the absorption spectra are quite similar for both the cases where absorption peaks are in the UV region. For common anion systems, the absorption edges are shifted towards the visible region with increasing atomic number of anions, while for common cation systems absorption edges are shifted from the visible to UV region when we go from Cd to Zn. So these systems may be useful for photovoltaic, light emitting diodes (LED) and  optoelectronic devices. We have also studied the electronic transport properties of the heterostructures and observed that both the systems give rise to asymmetric electronic transport, with the CdS-ZnS system showing more prominent effects in the calculated current-voltage characteristic. Such systems may have applications in electronic switching devices.

\section*{Acknowledgement}
We acknowledge the department of science and technology (DST) of India and Deutscher Akademischer Austauschdienst (DAAD) of Germany for funding. BD and ID thank TRC (DST) for computational support.

\bibliography{cqd_13Aug2022_final}
\end{document}